\documentclass[prd, twocolumn, nofootinbib]{revtex4}

\usepackage{epsfig} 

\newcommand{\beq}{\begin{equation}}
\newcommand{\eeq}{\end{equation}}
\newcommand{\beqa}{\begin{eqnarray}}
\newcommand{\eeqa}{\end{eqnarray}}
\newcommand{\om}{\Omega_m}
\newcommand{\omw}{\Omega_w}

\newcommand{\ls}{\mathrel{\raise0.27ex\hbox{$<$}\kern-0.70em \lower0.71ex\hbox{{
$\scriptstyle \sim$}}}}

\begin{document} 

%\title{Why {\lowercase {w}}=-1} 
\title{The Mirage of {\boldmath $w=-1$}} 
\author{Eric V. Linder}
%\affil{Berkeley Lab, University of California, Berkeley, CA 94720, USA} 
\affiliation{Berkeley Lab, University of California, Berkeley, CA 94720, USA} 
%\email{evlinder@lbl.gov}

%\date{\today}

\begin{abstract} 
We demonstrate that cosmic microwave background observations 
consistent with a cosmological constant universe predict in a well-defined 
sense that lower redshift measures will nearly automatically deliver 
$w=-1$ for the dark energy equation of state value unless they 
are sensitive to $w(z)$.  Thus low redshift data pointing to $w=-1$ does 
not truly argue for a cosmological constant. 
Even the simplest question of whether the equation of state of dark energy 
is equal to the cosmological constant therefore requires experiments able to 
sensitively constrain time variation $w(z)$ and not merely a constant $w$. 
We also note a number of issues regarding parametrization of $w(z)$, 
demonstrating that the standard form $w(z)=w_0+w_a 
z/(1+z)$ is robust but use of high order polynomials and cutting off the 
high redshift behavior can be pathological. 
\end{abstract} 

%\keywords{cosmology: observations --- cosmology: theory --- supernovae}
%\date{\today}

\maketitle

\section{Introduction \label{sec:intro}}

Advances in observational cosmology have led to tightening constraints 
on the properties of dark energy accelerating the cosmic expansion. 
Combinations of current data -- Type Ia supernovae distances (SN) together 
with cosmic microwave background (CMB) fluctuations and galaxy surveys 
measuring baryon acoustic oscillation (BAO) scales -- yield error bars 
on the effective pressure to energy density, or equation of state (EOS) 
ratio $w\approx-1\pm0.1$ (e.g.\ \cite{kowal}).  This represents 
significant progress in a first step toward checking consistency with 
a cosmological constant, possessing $w=-1$ at all redshifts. 

The generation of experiments now and in the next few years achieves 
constraints in terms of a time independent EOS, which can also be thought 
of as an average over the range of observations, a substantial fraction 
of a Hubble expansion time.  This article argues that such interpretation 
is very limited in providing new insight into dark energy, and will exhibit 
a definite tendency to deliver $w=-1$ despite real time variation. 
Therefore we should not consider experiments without specific sensitivity 
to time variation $w(z)$ as answering any questions about the nature of 
the accelerating universe.  Rather, they serve crucial roles as 
technological and methodological developments, in particular for controlling 
systematic uncertainties. 

Section \ref{sec:cmb} shows how the tendency arises that if CMB 
measurements of averaged dark energy influence implies $\langle w\rangle=-1$ 
then low redshift measurements will almost necessarily think $w=-1$. 
Since this does not imply that $w(z)=-1$, \S\ref{sec:w3} examines the 
robustness of parameterizing the high redshift behavior of $w(z)$. 
It is tempting to add more parameters to the EOS description, but 
\S\ref{sec:w4} points out several pitfalls and pathologies in this 
approach, as well as when it partially succeeds. 

The main purpose of this article is pedagogical emphasis on and 
clarification of the degeneracies and complementarity of different 
distance measures, and the essential need for experiments capable of 
precision measurements of the time variation of the equation of state, 
$w(z)$, before any real progress can be made in understanding the 
accelerating universe.

\section{Matching and Crossover \label{sec:cmb}}

Distances, whether luminosity distances of SN or angular distances of 
the CMB or BAO, involve integrals of the Hubble parameter and its 
constituent energy densities and double integrals of the dark energy 
equation of state.  This implies these quantities are intimately related 
and precise data in a distance will have implications for the allowed 
cosmological model parameters and density-redshift and EOS-redshift 
relations.  More explicitly, one can derive a 
chain of physical conditions, where matching distances 
in two models (so as to agree with data) leads to a convergence at 
certain redshifts in the behavior of the energy densities and a crossover 
(equal values) in the EOS.  

For example, \cite{francis} noted that matching the reduced distance 
to CMB last scattering $d_{\rm lss}$ led to a convergence of the dark 
energy density ratio between two models and a crossover 
in $w(z)=w_0+w_az/(1+z)$ at $a\approx0.7$.  Here we derive these relations 
analytically and investigate the implications for measurement of $w$. 

Suppose we fix the distance to some redshift $z$ to be some value, say 
$d_{\rm lss}$ to that of a concordance 
cosmological constant (LCDM) cosmology with matter density $\om=0.234$, 
in agreement with the analysis of \cite{spergel}.  If we also 
hold the present matter density fixed (for the moment, we address its 
variation later) then we can derive a quantity 
\beqa 
A&=&\int_1^Y dy\,\frac{\ln y}{(\om y^3+1-\om)^{3/2}}\nonumber \\  
&\quad&\Big/\int_1^Y dy\, 
\frac{\ln y+y^{-1}-1}{(\om y^3+1-\om)^{3/2}}, \label{eq:across} 
\eeqa 
where $Y=1+z$.  The dark energy density of the model, $\omw(z)$, will 
cross the curve of the fiducial dark energy density, e.g.\ 
$\Omega_\Lambda(z)$, at a crossover scale factor $a_\Omega$ solving 
\beq 
\frac{1-a_\Omega}{\ln a_\Omega}=A^{-1}-1. 
\eeq

Moreover, the dark energy EOS will cross $w=-1$ at a crossover scale 
factor 
\beq 
a_w=1-A^{-1}. \label{eq:aw}
\eeq 
We see there is a definite relation between $a_\Omega$ and $a_w$.  The 
$w=-1$ crossover is very robust.  If we change the fiducial value of 
$\om$ from 0.2 to 0.35 then $a_w$ ranges from 0.707 to 0.733, while 
$a_\Omega$ goes from 0.476 to 0.518.  

Thus a high redshift distance measurement 
consistent with LCDM virtually forces (within the picture so far) 
the value $w(z\approx0.4)=-1$, irrespective of true time variation. 
However, low redshift measurements insufficiently sensitive to time 
variation measure only an averaged EOS that corresponds strongly to 
the value at a sweet spot or ``pivot'' redshift with the pivot near 
$z\approx0.4$.  That is, the averaged EOS will necessarily appear to 
be $w\approx-1$, despite the presence of 
varying $w(z)$.  Not only is no new insight provided, but failing to 
recognize that the CMB and low redshift experiments are measuring 
essentially the same quantity could lead to the mistaken belief that 
strong evidence for $w=-1$ has been obtained and the nature of dark 
energy determined. 

\begin{figure}[!htb]
\begin{center} 
\psfig{file=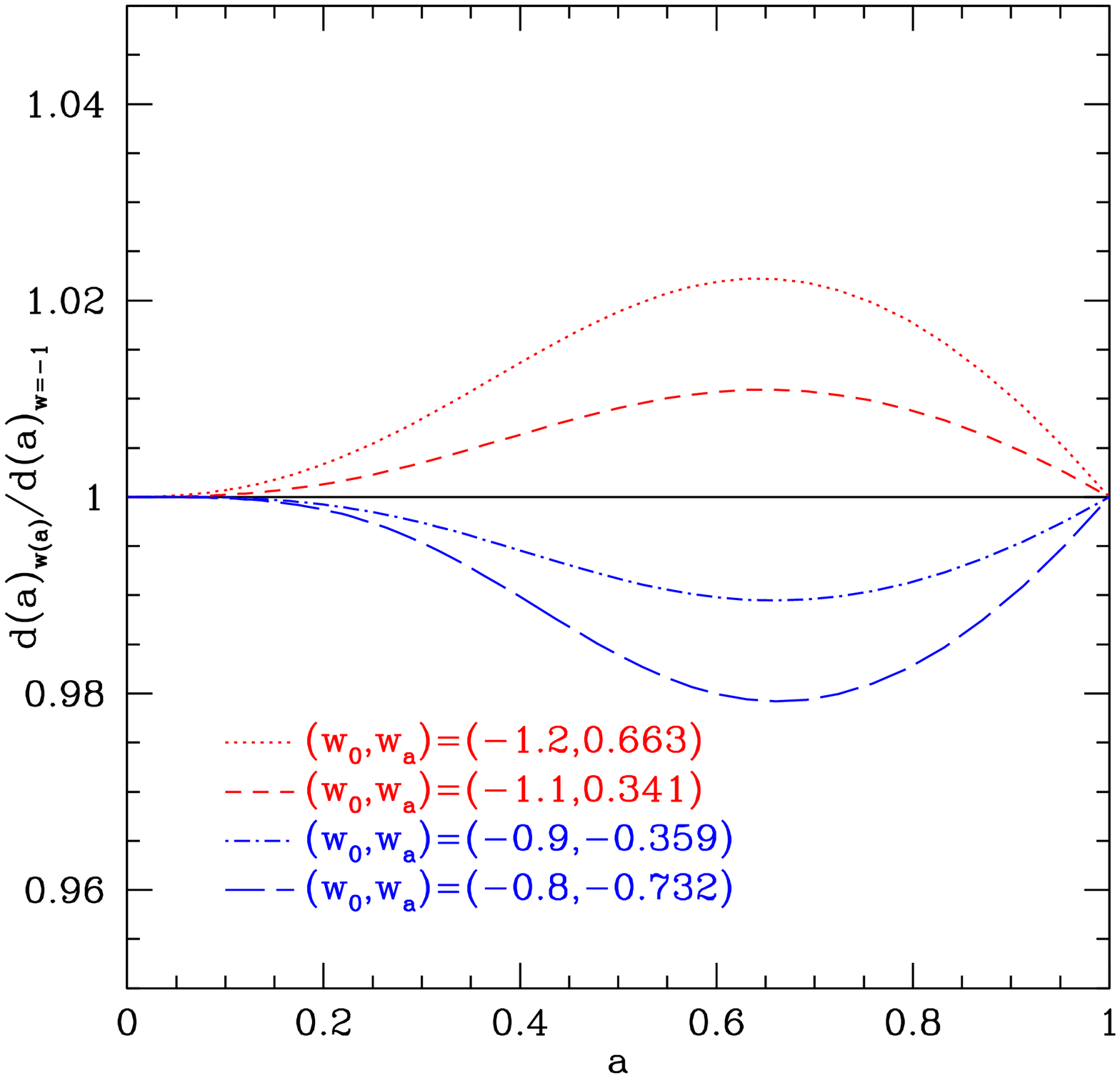,width=3.4in} 
\psfig{file=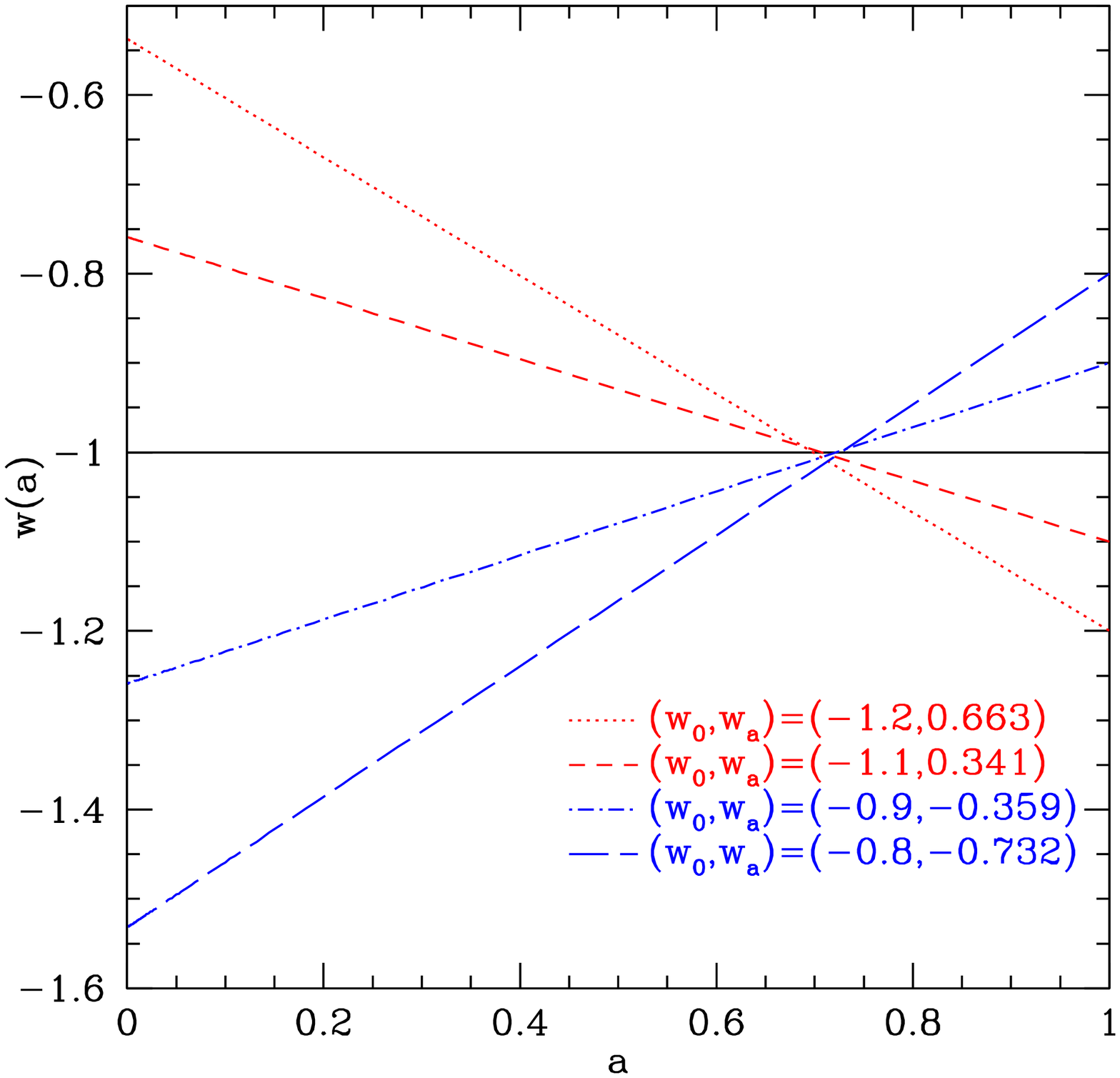,width=3.4in} 
\caption{Matching the distance to CMB last scattering between dark energy 
models leads to convergence and crossover behaviors in other cosmological 
quantities.  The top panel illustrates the convergence in the 
distance-redshift relation for models with $w_0$ ranging from $-0.8$ to 
$-1.2$ and corresponding $w_a$, relative to the $\Lambda$CDM case.  The 
bottom panel shows how this necessarily leads to a crossover with $w=-1$ 
at the key redshift for sensitivity of low redshift experiments.  The 
crossover in $w(z)$, and its uniqueness, is impelled by the physics not the 
functional form.  Here we fix $\om$; the text discusses robustness to 
allowing it to vary. 
}
\label{fig:xovers} 
\end{center} 
\end{figure}

Figure \ref{fig:xovers} demonstrates an example of how the CMB 
distance to last scattering forces low redshift experiments to 
measure $w=-1$.  While the crossover in $w$ between the LCDM fiducial 
case and the time varying models is not perfect, it is well within 
the precision of current and imminent experiments.  Such experiments 
insensitive to time variation essentially measure a constant $w$ given 
by its pivot value, hence $w\approx-1$.  The deviations in low redshift 
distances between the models are less than 2\%.  

\begin{figure}[!htb]
\begin{center} 
\psfig{file=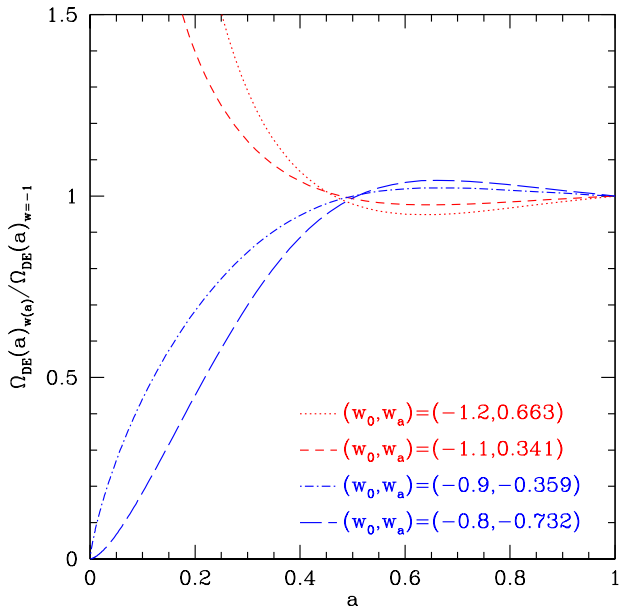,width=3.4in} 
\caption{As Fig.~\ref{fig:xovers}, showing how matching $d_{\rm lss}$ 
leads to convergence and crossover conditions on the dark energy 
density.  This has implications for growth of structure; see 
\cite{francis}. 
}
\label{fig:xovero} 
\end{center} 
\end{figure}

Conversely, note that 
Eq.~(\ref{eq:aw}) implies that to match LCDM for some model 
with current EOS $w_0$ one has $w_a=-A(1+w_0)$.  For $|1+w_0|<0.2$, 
the family of models generated in this way will all have the same 
$d_{\rm lss}$ to within 0.2\%, better than the anticipated accuracy of 
the Planck CMB experiment.  Also see Fig.~\ref{fig:xovero} where the 
matching carries through to $\om(z)$ and the growth of structure.

\begin{figure}[!htb]
\begin{center} 
\psfig{file=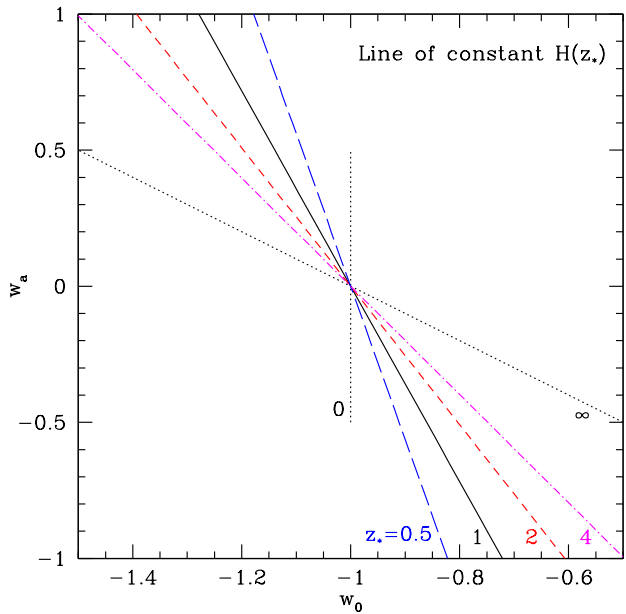,width=3.4in} 
\caption{Contours of constant expansion history $H(z)$ span a limited 
area of the dark energy equation of state plane $w_0$-$w_a$, even as 
redshift varies over $z=0-\infty$.  Here we hold $\om$ fixed so the lines 
also represent contours of constant $\om(z)$.  Any point along a given 
$z_*$ line represents a dark energy model with the same $H(z_*)$ as the 
$\Lambda$CDM model. 
}
\label{fig:omiso} 
\end{center} 
\end{figure}

One further implication of the matching-crossover relation for dark 
energy experiments is that there are 
unfortunately restrictions to the complementarity between different 
probes.  While techniques may probe the expansion history at various 
redshifts, hypothetically between $z=0$ and $\infty$, there 
remain degeneracies despite this range.  Measurements of the expansion 
history (and growth history) depend on the Hubble parameter $H(z)$, 
or matter density $\om(z)$.  Figure~\ref{fig:omiso} shows lines of 
constant $H(z)$ in the dark energy EOS parameter space.  We see that 
even for measurements at $z=0$ and at $z=\infty$ there is no true 
orthogonality.  This cannot be avoided for techniques involving the 
Hubble expansion in a simple way (including growth techniques, see 
\cite{coohut}).  Theoretically one can achieve full orthogonality, i.e.\ 
have a degeneracy direction running between upper right and lower left, 
by measuring distance ratios (e.g.\ via strong gravitational lensing 
\cite{linsl}) but this is undone by other degeneracies and systematic 
uncertainties.

Now let us relax the assumptions employed so far.  If we consider a family of 
models with present matter densities different from the fiducial, then 
the value of $w$ where the models cross upon matching to $d_{\rm lss}$, say, 
does differ from $-1$.  We find that the crossover remains at very close 
to the previous scale factor $a_w=0.7$ but the EOS value at the crossover 
adjusts to $w_p$ given by 
\beqa 
1+w_p&=&\frac{2}{3(\om^{-1}-1)}\Bigl[\int_1^Y dy\,\{[y^3+\om^{-1}-1]^{-1/2}
\nonumber\\ 
&\quad&\quad -[y^3+\bar\om^{-1}-1]^{-1/2}\}\Bigr]\nonumber\\ 
&\quad&\Big/\int_1^Y dy\,\ln y\,(y^3+\om^{-1}-1)^{-3/2},  
\eeqa 
where $\bar\om$ is the matter density in the fiducial LCDM model and 
$\om=\bar\om+\Delta\om$ is the model being considered.  This formula 
matches $d_{\rm lss}$ between models with $|\Delta\om|<0.1$ to 0.1\%. 
A good approximation is to take 
\beq 
1+w_p=3.6\,\Delta\om. 
\eeq 
(See also Eq.~(2) of \cite{fhlt}, and note we can apply our formula 
not just to constant $w$ but to the crossover value of the time varying 
EOS.) 

We can also relax the condition that the CMB measurements tell us 
precisely the distance to $z=1089$.  Planck data (temperature plus 
polarization) may provide a measurement of the reduced distance to last 
scattering of 0.4\%.  Allowing for errors in $d_{\rm lss}$, so models 
need not exactly match the LCDM fiducial best fit, introduces a scatter 
of 
\beq 
\sigma(w_p)\approx (+0.01,-0.02)\,\frac{|1+w_0|}{0.1}\,\frac{\delta 
d/d}{0.4\%}\,. 
\eeq 
The pivot scale factor stays near $a=0.7$. 

Thus, generalization of the CMB constraints to allow for uncertainty in 
the matter density and the last scattering distance measurement still 
preserves the implication that (so long as CMB data are consistent with 
a cosmological constant and -- a crucial point -- the uncertainty in 
the matter density is not too great) low redshift data will be driven to show 
$w\approx-1$. 
Despite this apparent confirmation of a cosmological constant, the dark 
energy models considered in Fig.~\ref{fig:xovers} actually have substantial 
true time variation, up to $|w_a|\approx1$.  Therefore measurements 
of $w=-1\pm0.05$, say, by ongoing and near term experiments {\it do not 
provide real support for a cosmological constant\/}.  Within the dark energy 
picture presented here, almost no answer other than $w=-1$ could have 
been expected.  Only future measurements directly sensitive to time 
variation can truly add to knowledge of whether $w=-1$ or not.

\section{Robust Parametrization \label{sec:w3}}

One possible loophole in the picture presented is if the dark energy 
behavior differs substantially from the parametrization for $w(z)$ we 
adopted.  However we emphasize that the 
form $w(z)=w_0+w_az/(1+z)$ does not force models to cross and 
certainly does not force them all to cross at the same scale factor.  
They are impelled to do so by the physics not the form.  
Nevertheless, we want to ensure that this parametrization is a robust 
description of a wide range of dark energy behaviors, especially at high 
redshift.  This has been addressed in a number of articles, e.g.\ 
\cite{linprl,linbias}, but here we concentrate on testing high redshift and 
deviations from a linear dependence on scale factor $a$. 

One model specifically designed to consider dark energy influence at 
high redshift is the bending model of \cite{wett}, with 
\beq 
w(z)=w_0\,[1+b\ln(1+z)]^{-2}. \label{eq:wbend}
\eeq 
The bending parameter $b$ is directly related to the early dark energy 
density $\Omega_e=\Omega_w(z\gg1)$.  We consider $w_0=-0.9$ and $b=0.415$, 
corresponding to $\Omega_e=0.02$, close to the maximum allowed by data. 
Note that in contrast $\Omega_\Lambda(z=1089)\approx 10^{-9}$.  While 
the $w_0$-$w_a$ model was not designed to fit early dark energy, it does 
an admirable job.  The model $(w_0,w_a)=(-0.9,0.7)$ can match SN distances 
in the bending 
model to better than 0.004 magnitudes out to $z=2$ and agrees on $w(z)$ 
out to $z=2$ to 2\%, while $d_{\rm lss}$ is within 0.4\%.  

\begin{figure}[!htb]
\begin{center} 
\psfig{file=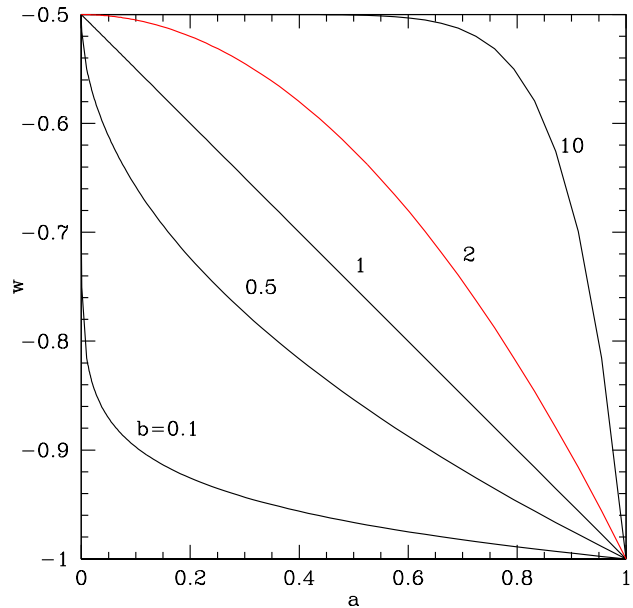,width=3.4in} 
\caption{Equation of state as a function of scale factor for the form 
$w(a)=w_0+w_a(1-a^b)$, with curves labeled by $b$. Increasing $b$ gives 
more rapid time variation recently.  The standard $w_0$-$w_a$ form 
provides a robust fit for equation of state and distances for all 
cases $b\lesssim2$, corresponding to $w_a\lesssim1$. 
}
\label{fig:wab} 
\end{center} 
\end{figure}

Suppose we want to investigate dark energy with a more rapid evolution 
than apparent from the $w_a$ model, and are willing to include a third 
parameter.  Consider model 3.1 of \cite{lh05}, 
\beq 
w(a)=w_0+w_a(1-a^b), \label{eq:wab}
\eeq 
where the usual $w_a$ model corresponds to $b=1$.  Figure~\ref{fig:wab} 
shows the diverse time variation as we adjust $b$.  Despite this, the 
standard $w_a$ model can successfully fit variations that are not too 
extreme, $b\lesssim2$, to good accuracy.  Taking a fairly substantial 
time evolution, from $w_0=-1$ to $w(z\gg1)=-0.5$, we find that 
the pure $w_a$ model fits $w(z)$ to 0.6\%, 1.6\% (2\%, 4.5\%) accuracy 
compared to the cases with $b=0.5$, 1.5 at $z=0.5$ ($z=2$), and SN 
distances to 0.005 mag out to $z=2$ and $d_{\rm lss}$ to 0.07\%.  
Only very rapid variations, with $b>2$, corresponding to $dw/dz>1$ today, 
will cause problems for the standard $w_a$ parametrization.

\begin{figure}[!htb]
\begin{center} 
\psfig{file=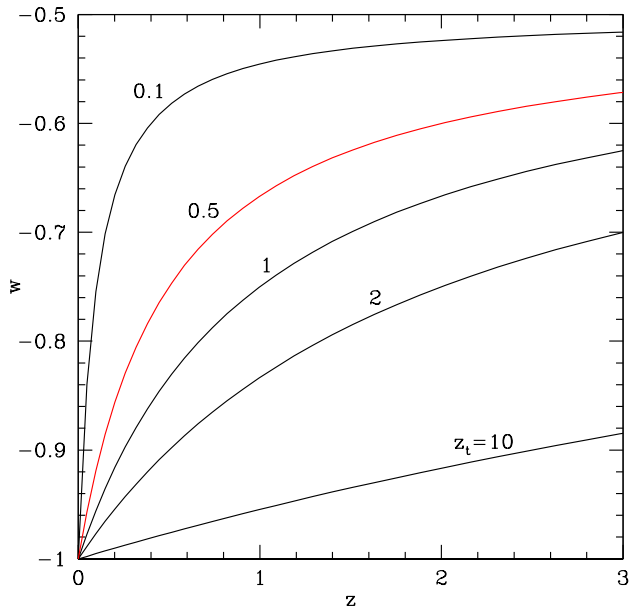,width=3.4in} 
\caption{Equation of state as a function of redshift for the form 
$w(z)=(w_\infty z+w_0z_t)/(z+z_t)$, with curves labeled by $z_t$.  
Decreasing $z_t$ gives more rapid time variation recently.  The standard 
$w_0$-$w_a$ form 
provides a robust fit for equation of state and distances for all 
cases $z_t\gtrsim0.5$, corresponding to $w_a\lesssim1$. 
}
\label{fig:wext} 
\end{center} 
\end{figure}

One could also alter the time characterizing when the EOS variation 
occurs, with a third parameter as in the extension model of \cite{rapaw}, 
\beq 
w(z)=\frac{w_\infty z+w_0z_t}{z+z_t}, \label{eq:wext} 
\eeq 
where $w_\infty=w(z\gg1)$ and this form reduces to the $w_a$ model for 
a transition redshift $z_t=1$.  The impact of such variation on the EOS 
is shown in Fig.~\ref{fig:wext}, with smaller $z_t$ causing a more rapid 
transition in the recent universe.  Again, the usual $w_a$ model can 
encompass this behavior, 
reproducing $w(z)$ even in the substantially time varying case of $w_0=-1$, 
$w_\infty=-0.5$ to within 1\%, 4\% (4\%, 8\%) for $z_t=2$, 0.5 at $z=0.5$ 
($z=2$).  Distances agree to within 0.013 mag out to $z=2$ and to 0.002\% 
for $d_{\rm lss}$.  
Only very rapid variations, with $z_t<0.5$, corresponding to $dw/dz>1$ 
today, will cause problems for the standard $w_a$ parametrization.

Note that in none of the three forms we have considered have we applied 
an optimization for the $w_a$ fit utilizing the distance matching-crossover 
physics discussed in \S\ref{sec:cmb}, i.e.\ we have not deliberately 
matched distances.  These cases demonstrate that the 
standard $w_0$-$w_a$ parametrization has substantial robustness, even for 
fairly rapidly varying EOS behavior or dark energy with significant presence 
at high redshifts.  The $w_a$ robustness is not perfect, of course; very 
rapid variation, with $w_a>1$, or significant nonmonotonic behavior 
in $w(z)$ will cause the parametrization to break down.  However, for a 
large variety of behavior, and in particular behavior where an averaged 
EOS $w$ might be thought to hold insight, the $w_a$ parametrization is robust 
and the conclusions of \S\ref{sec:cmb} regarding the appearance of $w=-1$ 
are not in jeopardy. 

\begin{figure}[!htb]
\begin{center}
\psfig{file=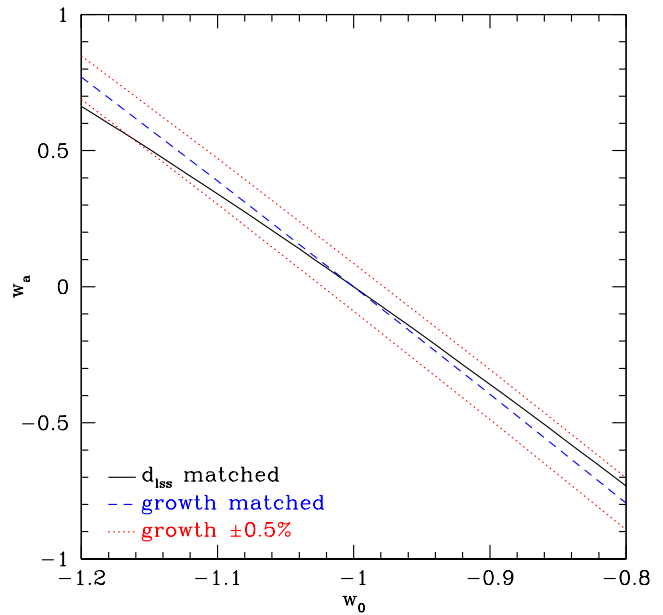,width=3.4in}
\caption{Matching the CMB distance to last scattering is closely 
related to matching the particular linear growth factor ratio 
$g(a=0.35)/g(a=1)$ of \cite{linwhite}.  Here we show the similarity of 
degeneracy 
directions in the dark energy equation of state plane, with $\om=0.234$. 
}
\label{fig:growcmb}
\end{center}
\end{figure}

It is interesting to go one step further, and consider the growth history 
as an observable.  Recall that \cite{linwhite,francis} showed there exists 
(within general relativity) a close relation between the distance to CMB 
last scattering and the linear growth factor of density perturbations. 
We illustrate this relation in Fig.~\ref{fig:growcmb} for the ratio of 
growth factors $g(a=0.35)/g(a=1)$ identified in \cite{linwhite}. 
We can also ask what a measurement of the absolute linear growth factor at, 
say, $z=2$ can teach us about the high redshift dark energy behavior.  Fixing 
the characteristics of the low redshift universe, say $\om=0.28$ and 
$w_0=-1$, if the growth at $z=2$ agrees with the LCDM model to 5\% 
then within the $w_a$ model (which, recall, works well for early dark 
energy) we can limit $w_a<0.6$ or $w(z=2)<-0.6$.

\section{Problems in Parametrization \label{sec:w4}}

Despite the success of the $w_0$-$w_a$ model, one might be tempted to 
extend it to more parameters, such as the three parameter models discussed 
in \S\ref{sec:w3}, or more general high order polynomial fits.  However, 
the number of EOS parameters that can be accurately fit by even the 
combination of next generation data has been shown to be limited to 
two \cite{lh05}.  Nevertheless, let us explore this slightly further. 

One might wish to consider more parameters and simply marginalize over 
the extra parameters beyond two.  If this works, i.e.\ does not strongly 
degrade the main two parameters' estimation, then this effectively allows 
more freedom in the functional form.  

Consider the ``ab'' form of 
Eq.~(\ref{eq:wab}).  Next generation SN+CMB data cannot fit the third 
parameter and including $b$ weakens estimation of $w_0$ and $w_a$ by 
more than an 
order of magnitude.  Further addition of next generation weak lensing 
data does not help.  However, we find that if we place a prior on the 
third parameter, then while the third parameter is not better determined 
than our input prior, the first two EOS parameters are not strongly 
degraded when the prior satisfies 
\beq 
\sigma_b<\frac{1}{3\sim10}\frac{1}{w_a}. \label{eq:priorab}
\eeq 
For the coefficient $1/3$ ($1/10$), the degradation in $w_a$ is 15-20\% 
($\sim$2\%), in $w_0$ is less than 1\%, and in the always more unstable 
pivot value $w_p$ is 60-130\% (10-30\%).  When $w_a<0$ then the required 
prior is weaker, with the $1/3$ coefficient acting as strongly as the 
$1/10$ coefficient does for positive $w_a$. 

The main effect of a constrained third, marginalized parameter is therefore 
a thickening of the confidence contour (greater covariance) rather than 
an elongation of the confidence interval of the parameter values.  
In this sense, 
it acts substantially like a systematic uncertainty, but here the 
prior or limit on the uncertainty, e.g.\ $\sigma_b$, is likely arbitrary.  
So exploration of a third parameter can enlarge the freedom of the 
functional form of $w(z)$, but within arbitrary limits.  It also weakens 
the main parameter estimation.  That penalty will be most severe when 
considering ``figures of merit'' based on the area $\sigma(w_p)\cdot 
\sigma(w_a)$ of the confidence region, but relatively benign if the 
main physical questions involve $\sigma(w_0)$ or $\sigma(w_a)$ as 
discussed by \cite{linbias}. 

To check these results, we also considered the $w_{\rm ext}$ form of 
Eq.~(\ref{eq:wext}).  This leads to very similar conclusions, that to 
avoid degradation of the main two EOS parameters one needs a 
prior on the third parameter $z_t$ of order 
\beq 
\sigma_{z_t}<\frac{1}{3w_a}, \label{eq:priorext}
\eeq 
where $w_a=w_\infty-w_0$.  Note that \cite{rapaw} considered 
marginalization over $z_t$ with a prior of 0.5 and $w_a\approx0.6$. 
In both the ab and ext cases one can understand the required magnitude 
of the prior, Eqs.~(\ref{eq:priorab})-(\ref{eq:priorext}), by studying 
the form of the Hubble parameter. 

While these instances of adding EOS parameters beyond two are fairly 
benign under the given conditions, using many parameter forms can be 
not only useless (since only two parameters can be fruitfully 
constrained) but dangerous.  For example, consider a general fourth 
order polynomial 
\beq 
w(z)=w_0+\sum_{i=1}^4 w_i\,[\ln(1+z)]^i, \label{eq:poly} 
\eeq 
as discussed in \cite{riess06}.  Attempting to fit five EOS parameters 
does not yield useful information even with the combination of next 
generation data, as mentioned above.  However, its implementation does 
create pathologies, biasing the results. 

Consider the bending case of Eq.~(\ref{eq:wbend}) that is adept at 
describing early dark energy.  We saw that the $w_a$ parametrization 
could match this to high accuracy.  However, the polynomial form of 
Eq.~(\ref{eq:poly}) is unbounded at high redshift and so must be cut off. 
In \cite{riess06} and in its use for projection of a next generation 
dark energy mission, the convention is to fix $w=-1$ for $z>2$. 
Let us examine the consequences of this. 

\begin{figure}[!htb]
\begin{center} 
\psfig{file=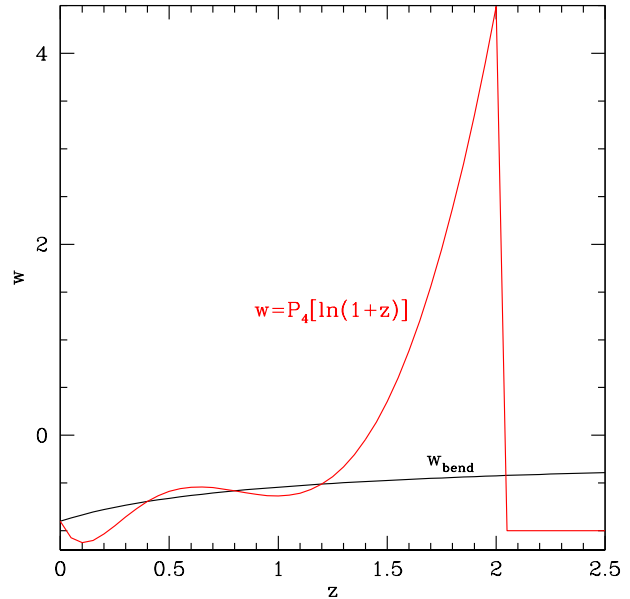,width=3.4in} 
\caption{The fourth order polynomial form (\ref{eq:poly}) for $w(z)$ 
exhibits pathological behavior and poor fit to equation of state and 
distances even when fixed to the true $w(z)$ at four redshifts and matched 
to give the correct $d_{\rm lss}$.  Here the true model is the bending 
form (\ref{eq:wbend}) for early dark energy. 
}
\label{fig:p4} 
\end{center} 
\end{figure}

Suppose we could tune the fourth order polynomial to match exactly at $z\le2$ 
the EOS in the bending case of $w_0=-0.9$, $b=0.415$ considered in 
\S\ref{sec:w3}.  Then due to the cutoff it will misestimate the distance 
to CMB last 
scattering by 2.2\%, greater than not only Planck precision but current 
WMAP precision.  Instead let us adjust the polynomial coefficients to 
match $d_{\rm lss}$ and $w(z)$ at $z=0$, 0.4, 0.8, and 1.2.  One might 
think that this would provide excellent approximation to the EOS and 
the distances -- after all, one is matching many more quantities than 
the mere two of $w_0$, $w_a$.  However, the result, 
shown in Fig.~\ref{fig:p4}, is that the fourth order polynomial
wildly oscillates.  This form misestimates SN distances by 0.075 mag 
(recall that $w_a$ succeeded to within 0.004 mag, almost 20 times better).

\begin{figure}[!htb]
\begin{center} 
\psfig{file=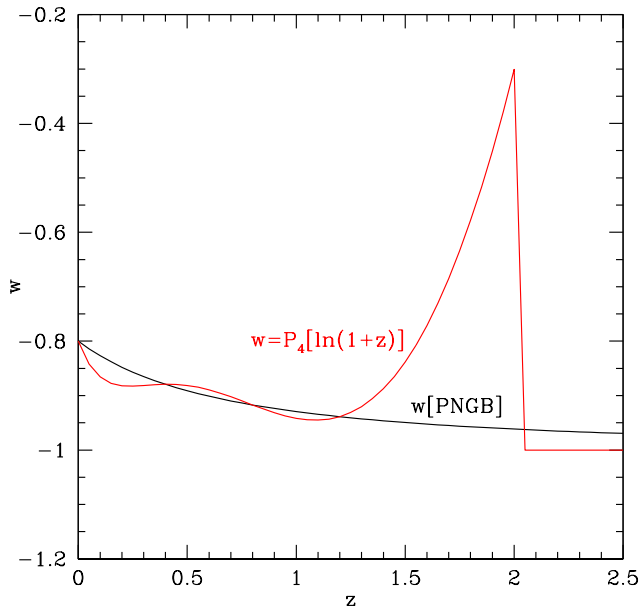,width=3.4in} 
\caption{As Fig.~\ref{fig:p4}, but for the PNGB thawing dark energy 
model, showing that the pathology in the fourth order polynomial 
persists even for models that approach $w=-1$ at high redshift. 
}
\label{fig:p4thaw} 
\end{center} 
\end{figure}

This pathology can be traced to the cutoff at $z>2$.  As with spline fits, 
the freedom in the form often leads to spurious wiggles, especially if the 
boundary conditions are not carefully chosen.  
Even thawing dark energy models, which indeed approach $w(z\gg1)=-1$, 
are not well fit by the fourth order polynomial form.  Taking a simple 
PNGB dark energy model $w(z)=-1+(1+w_0)(1+z)^{-F}$ with $w_0=-0.8$ and 
$F=1.5$ yields the results in Fig.~\ref{fig:p4thaw}.  In fact, in 
seeking a nonpathological fit one is driven to $w_2=w_3=w_4=0$, back to 
a two parameter form such as $w_0$, $w_a$ so successfully demonstrates.

\begin{figure}[!htb]
\begin{center}
\psfig{file=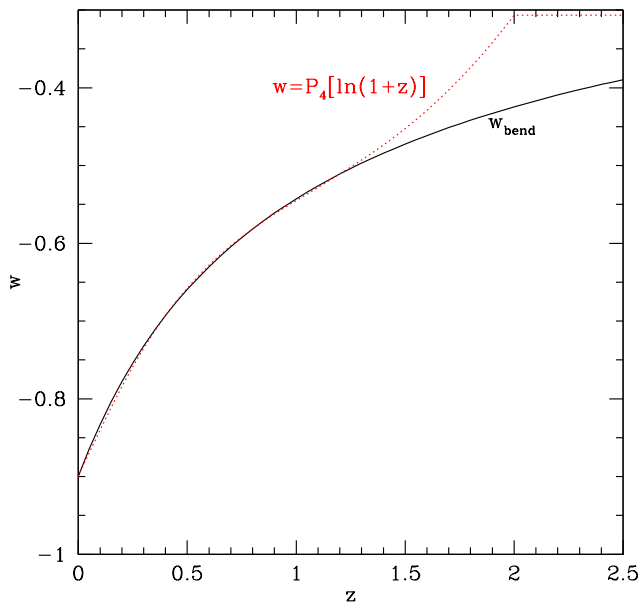,width=3.4in}
\caption{As Fig.~\ref{fig:p4}, but we remove the condition $w(z>2)=-1$ 
and instead use $w(z>2)=w(z=2)$.  This tames the spurious oscillations 
considerably.  However, note that the form (\ref{eq:poly}) using five 
parameters still has difficulties fitting a two parameter equation of 
state model. 
}
\label{fig:p4c}
\end{center}
\end{figure}

One can ameliorate considerably the problems of the fourth order 
polynomial by instead choosing $w(z>2)=w(z=2)$, or more generally 
\beq 
w(z>z_{\rm cut})=w(z=z_{\rm cut}). 
\eeq 
This delivers much better results: the distances out to $z=2$ in the 
bending case are now within 0.002 mag of the true values (also see 
Fig.~\ref{fig:p4c}).  However, 
this merely fixed a problem that should never have arisen: the data 
can actually fit no more than two EOS parameters so there is no point 
in considering high order polynomials.

\section{Conclusion \label{sec:concl}}

Given accurate measurement of the distance to CMB last scattering 
consistent with LCDM, 
interpreting low redshift distance data leading to $w=-1$ as evidence 
for a cosmological constant considerably overstates the case.  By the 
nature of $z\lesssim1.5$ observations with insufficient accuracy to 
recognize time variation in the equation of state, they essentially 
measure an averaged $w$ or equivalent constant $w$ just where the CMB 
data and the cosmological relation between distance and EOS ineluctably 
predict that constant $w=-1$ to a few percent accuracy\footnote{Since 
reasonable bounds on the matter density $\om$ play an important role 
in this, this highlights the value of obtaining accurate measurements 
of the Hubble constant to combine with constraints on $\om h^2$, as 
\cite{hu} has so eloquently argued.}. 

Even models with time variation $w_a\approx1$ can appear to show $w=-1$ 
when viewed by these experiments.  Such measurements of $w\approx-1$ 
induce a false sense of security in $\Lambda$.  To achieve true 
insight into the 
nature of dark energy, even as to simply whether or not it is a cosmological 
constant, requires measurements capable of directly probing the time 
variation with significant sensitivity.  Current and near term experiments 
serve a valuable and necessary role in developing techniques and tightening 
controls on systematic uncertainties. 

These conclusions arise within a parametrization of the dark energy 
behavior, at least over the range where it significantly affects the 
distance-redshift relation, of $w(z)=w_0+w_az/(1+z)$.  Checking this 
assumption, we find this form provides an excellent, robust approximation 
to a wide variety of behaviors, even fairly rapidly evolving ones 
$w_a\lesssim1$ (though 
we have not considered nonmonotonic behaviors, which are not expected 
from single field models traced over the age of the universe, and as 
seen can easily be spurious, from noise or improper fits). 

Apart from the issue of $w=-1$, attempting to fit the equation 
of state with more than two parameters is not only fruitless (if significant 
accuracy is desired) but can lead to pathologies.  We have demonstrated 
this in the case of a proposed fourth order polynomial.  One can also attempt 
to marginalize over a third parameter, to expand the functional freedom, 
but it is unclear what this gains (since $w_0$-$w_a$ does a good job 
fitting these forms on its own) and it requires an arbitrary prior on 
the third parameter to prevent bloating of the confidence regions of the 
other parameters. 

Even for the basic question of whether the dark energy is Einstein's 
cosmological constant, there do not appear to be any short cuts before 
a carefully designed experiment to constrain accurately the time evolution 
of the dark energy equation of state, $w(z)$ not merely $w$. 

\acknowledgments

I thank the Aspen Center for Physics and Santa Fe Cosmology workshop SF07 
for hospitality, and the participants in SF07, especially Neal Dalal, for 
useful questions.  I also thank my earlier collaborators Matthew Francis, 
Geraint Lewis, and Martin White for motivating thoughts on the 
matching/crossover implications.  This work has been supported in part by 
the Director, Office of Science, 
Department of Energy under grant DE-AC02-05CH11231.

\end{document}